\documentclass[11pt,a4paper]{article}
 
\usepackage{amssymb} 
\usepackage{amsmath} 
\usepackage[]{graphicx} 
 
\newcommand{\be}{\begin{equation}} 
\newcommand{\ee}{\end{equation}} 
\newcommand{\bea}{\begin{eqnarray}} 
\newcommand{\eea}{\end{eqnarray}} 
\newcommand{\lb}{\label} 
\newcommand{\bdm}{\begin{displaymath}} 
\newcommand{\edm}{\end{displaymath}} 
\newcommand{\D}{{\rm d}} 
\newcommand{\md}{{\mathrm{d}}} 
 
\newcommand{\I}{{\rm i}} 
 
\def\cF{{\mathcal F}} 
\def\cO{{\mathcal O}} 
 
\begin{document} 
 
\thispagestyle{empty} 
 
\noindent 
\begin{center} 
\vspace*{1cm} 
 
{\large\bf QUANTUM GRAVITATIONAL COLLAPSE IN THE 
  LEMAITRE--TOLMAN--BONDI MODEL WITH A POSITIVE COSMOLOGICAL CONSTANT} 
 
\vskip 1cm 
 
\vskip 0.3cm

{\bf Anne Franzen}\footnote{New address from June 1, 2009: Institute 
  for Theoretical Physics, Utrecht University, Leuvenlaan 4, 3584 CE 
  Utrecht, The Netherlands}  
\vskip 0.3cm 
Institute for Theoretical Physics,\\ University of Cologne, \\ 
Z\"ulpicher Strasse~77, 50937 K\"oln, Germany 
 
\vskip 6mm 
 
{\bf Sashideep Gutti}\footnote {Part of the work was completed before
  August 2008 at Tata Institute of Fundamental Research, Homi Bhabha
  Road, Mumbai, 400005, India.}  
\vskip 0.3cm 
The Harish-Chandra Research Institute,\\ 
Chhatnag Road, Jhunsi, Allahabad, \\ Uttar Pradesh 
211019, India 
 
\vskip 6mm 
 
{\bf Claus Kiefer} 
\vskip 0.3cm 
Institute for Theoretical Physics,\\ University of Cologne, \\ 
Z\"ulpicher Strasse~77, 
50937 K\"oln, Germany. 
\vspace{1cm} 
 
\begin{abstract} 
Previous papers dealt with the quantization of the 
Lema\^{\i}tre--Tol\-man--Bondi (LTB) model for vanishing cosmological 
constant $\Lambda$. Here we extend the analysis to the case $\Lambda >0$. 
Our main goal is to present solutions of the Wheeler--DeWitt equation, 
to give their interpretation, and to derive Hawking radiation from 
them. We restrict ourselves to a discussion of those points that are 
different from the $\Lambda=0$-case. These have mainly to do with the 
occurrence of two horizons. 
\end{abstract}

\end{center}

\newpage 
\setcounter{page}{1} 

\section{Introduction}
 
Part of the research on quantum gravity consists in the attempt to get
a possible insight into the final theory from concrete models.
The simplest models deal with spatially
homogeneous cosmological metrics and are typically used in quantum 
cosmology. Less simple are models with spherical symmetry. They are 
used in cosmology, too, but mainly in the description of black holes. 
 
Here we deal with a particular spherically-symmetric model, the 
Le\-ma\^{\i}\-tre--Tol\-man--Bondi (LTB) model. It describes 
self-gravitating inhomogeneous dust and is well understood at the 
classical level where it is mostly used in cosmology \cite{PK}. The 
quantization of this model was 
attempted in both quantum geometrodynamics and loop quantum 
gravity. Here we restrict ourselves to quantum geometrodynamics, in 
which the central equations are the Wheeler--DeWitt equation and the 
momentum (diffeomorphism) constraints \cite{OUP,GRG}. 
 
Following the  application of canonical geometrodynamics to the 
Schwarz\-schild black hole \cite{Kuchar} and to 
Reissner--Nordstr\"om--anti-de~Sitter black holes \cite{LWH}, 
the canonical formalism for the LTB model was developed in 
\cite{VWS01}  and 
then applied to quantization in a series of papers, see 
\cite{KMHV06,KMHSV07,VGKS07} and the references therein. While it was not 
possible to construct the quantum theory for the LTB 
model in a rigorous way, partial progress was achieved. Among the 
results was the 
recovery of Hawking radiation plus greybody corrections from exact 
solutions to the Wheeler--DeWitt equation and the momentum 
constraints \cite{KMHSV07,VGKS07}. 
They were found for the case of vanishing 
cosmological constant $\Lambda$ where only one horizon (the black-hole 
horizon) is present. Here we extend the analysis to the case 
$\Lambda>0$, where two horizons (the black-hole and the cosmological 
horizon) exist. We again manage to find exact quantum states 
from which the two different Hawking temperatures from the two 
horizons can be recovered. 

Although we are mainly interested in understanding quantum gravity, we
want to emphasize that the case $\Lambda>0$ fits very well current
cosmological observations \cite{Hinshaw}. The analysis for
$\Lambda<0$, where only one horizon is present, was performed in
\cite{VTS08}. 
 
In our present paper we shall focus on the main differences to the 
case of vanishing $\Lambda$. For more details on the general formalism 
we refer to the earlier papers cited above. The LTB model was also 
addressed from the perspective of loop quantum gravity
\cite{ltbloop}. 

\section{The classical LTB model with positive cosmological constant}
 
Since we deal with dust, the energy--momentum tensor is given by $T_{\mu 
\nu}= \epsilon(\tau,\rho)u_\mu u_\nu$, where $u^\mu=u^\mu 
(\tau,\rho)$ is the four-velocity vector of a dust particle with 
proper time $\tau$; the parameter $\rho$ labels the various shells 
which together form the dust cloud. 
The line element for the LTB spacetime in comoving synchronous 
coordinates is given by 
 
\begin{align} 
\label{ltb-metric} 
\mathrm{d}s^2 &= -\mathrm{d}\tau^2 + 
\frac{(\partial_{\rho}R)^2}{1+2E(\rho)} \mathrm{d}\rho^2 
+ R^2(\rho,\tau)(\mathrm{d}\theta^2 + \sin^2\theta \mathrm{d}\phi^2)\ . 
\end{align} 
Inserting this expression into the Einstein field equations leads to 
\be \label{ltb-eg1} 8\pi G\epsilon(\tau,\rho) = 
\frac{\partial_{\rho}F}{R^2 \partial_{\rho}R} 
\ee 
and 
\be \label{ltb-eg2} 
  (\partial_{\tau}R)^2 = \frac{F}{R} 
  +\frac{\Lambda R^2}{3}+2E\equiv 1-{\mathcal F}+2E\ , 
\ee 
where $F(\rho)\equiv 2GM(\rho)$ is a non-negative function with the 
dimension of a length, and 
\be\lb{mathcalF} 
{\mathcal F}\equiv 1-\frac{F}{R}-\frac{\Lambda R^2}{3}\ . 
\ee 
The case of collapse is described by $\partial_{\tau}R(\tau,\rho)<0$. 
We set $c=1$ throughout. 
 
There still exists the freedom to rescale the shell index $\rho$. 
This freedom can be removed by demanding 
\begin{equation} \label{ltb-fix-r} R(0,\rho) = \rho \, , \end{equation} 
so that for $\tau=0$ the label coordinate 
$\rho$ is equal to the curvature radius $R$. Now we can express the functions 
$F(\rho)$ and $E(\rho)$ in terms of the energy density $\epsilon$ at 
$\tau=0$. From \eqref{ltb-eg1} and \eqref{ltb-eg2} one gets 
\begin{align} \label{ltb_F} F(\rho) &=8\pi G 
\int_0^{\rho} \epsilon(0,\tilde \rho)\, 
\tilde{\rho}^2\,\mathrm{d}\tilde \rho\ , \\ E(\rho) &=\frac12 
[\partial_{\tau}R(\tau=0,\rho)]^2-\frac{F(\rho)}{2\rho} 
-\frac{\Lambda \rho^2}{6}\; .\end{align} 
The interpretation of these quantities is that 
$F(\rho)/2G\equiv M(\rho)$ is the active gravitating mass within a 
$\rho=$ constant shell, 
while $E(\rho)$ is the total energy per unit mass within the same shell. 
The marginally bound models are defined by $E(\rho)\equiv 0$. 
Here we consider the general case, which includes 
the non-marginal case defined by $E(\rho)\neq0$. 
 
In order to derive Hawking radiation, we shall consider the following 
situation. We assume the presence of a black hole with mass $M_0$ 
surrounded by a 
gravitating dust cloud. We assume, moreover, that the total energy of 
the cloud is small compared to the mass of the black hole. 
After quantization, we shall find exact 
quantum states for the dust cloud from which the Hawking temperatures 
will be found. The dust cloud will play the role of the quantum fields 
usually employed in the derivation of Hawking radiation. 
 
If the dust cloud is of finite extension, 
the metric outside of it is of the Schwarzschild--de~Sitter (SdS) or Kottler 
form,\footnote{See, for example, \cite{Geyer} for a detailed 
  discussion of this metric.} 
\bea 
\lb{sdsmetric} 
\D s^2&=&-\left(1-\frac{2GM}{R}-\frac{\Lambda R^2}{3}\right)\D 
T^2\nonumber\\ & & \; +\left(1-\frac{2GM}{R}-\frac{\Lambda 
    R^2}{3}\right)^{-1}\D R^2 +R^2\D \Omega^2, 
\eea 
where $M$ is the mass of the black hole plus the mass contribution of 
the cloud. 
For small $R$, SdS space approximates Schwarzschild space, while for 
large $R$ and positive $\Lambda$ it approximates de Sitter space. 
The black-hole horizon $R_{\rm h}$ and the cosmological horizon 
$R_{\rm c}$ are two of the three zeros of $g_{TT}$ in 
\eqref{sdsmetric}, see, for example, \cite{GH77} and 
\cite{Stotyn}. They are explicitly given by 
\bea 
\lb{horizons} 
R_{\rm h} &=& 
3GM\ell\xi\left(1-\sqrt{1-\frac{1}{\ell\xi^3}}\right)
=\frac{\xi}{\sqrt{\Lambda}}\left(1-\sqrt{1-\frac{1}{\ell\xi^3}}\right)\ ,
\nonumber\\ 
R_{\rm c} &=& 
3GM\ell\xi\left(1+\sqrt{1-\frac{1}{\ell\xi^3}}\right)
=\frac{\xi}{\sqrt{\Lambda}}\left(1+\sqrt{1-\frac{1}{\ell\xi^3}}\right)
\ , 
\eea 
where $\ell^{-1}=3GM\sqrt{\Lambda}$ and 
$\xi=\cos(\frac13\cos^{-1}[\ell^{-1}])$. 
The third zero value, $R_{\rm n}$, is negative, 
$R_{\rm n}=-R_{\rm h}-R_{\rm c}=-6GM\ell\xi$ and possesses no 
obvious meaning.\footnote{It must be assumed here that the 
black-hole horizon is smaller than the cosmological horizon, $R_{\rm 
  h}<R_{\rm c}$, which means that $\ell^{-1}<1$. This is well 
satisfied in our Universe: Inserting for $M$ the mass of the 
supermassive black hole in the quasar OJ287, which is presently the 
biggest known supermassive black hole, with $M\approx 1.8\times 10^{10} 
M_{\odot}$ \cite{Valtonen}, and for $\Lambda$ the value following from 
$\Omega_{\Lambda}\approx 0.726$ \cite{Hinshaw}, we get 
$\ell^{-1}\approx 9\times 10^{-11}$. The value $\ell=1$ would 
correspond to the Nariai metric for which both areas are the same but are 
separated by a finite proper distance. (See \cite{Krasinski} for a 
discussion of the Nariai metric and a reprint of Nariai's original 
papers.)}  
The LTB solution \eqref{ltb-metric} must then be matched at the dust
boundary to the SdS solution.
 
In the canonical formalism, the general ansatz for a 
spherically-symmetric line element reads 
\begin{align}\lb{metric}\mathrm{d}s^2 
  = -N^2 \mathrm{d}t^2 + L^2 \left( \mathrm{d}r - N^r \mathrm{d}t \right)^2 + 
  R^2 \mathrm{d} \Omega^2 \ ,  \end{align} 
where 
$N$ is the lapse function, $N^r$ is the only component of the shift 
vector that survives the symmetry reduction, and $\mathrm{d} 
\Omega^2=\D \theta^2 + \sin^2 \theta \D \phi^2$ is the line element 
on the unit two-sphere. 
 
Inserting the ansatz \eqref{metric} into the ADM form of the 
Einstein--Hilbert action, we obtain the gravitational part of the 
action, 
\begin{align} S^{\mathrm{g}} = \int 
  \mathrm{d}t \int_0^{\infty} \mathrm{d}r \left(P_L \dot{L} + P_R \dot{R} - N 
    {\mathcal H}^{\rm g} - N^r {\mathcal H}_r^{\rm g} \right) + 
S_{\partial \Sigma} \; , 
\label{mega_action}\end{align} 
where the Hamiltonian and the diffeomorphism (momentum) constraint are 
given by 
\begin{align} {\mathcal H}^{\rm g} &= - G\left(\frac{P_L P_R}{R} - 
    \frac{LP_L^2}{2 R^2}\right) + 
  \frac{1}{G}\left[ -\frac{L}{2} - \frac{R'^2}{2L}+ 
    \left(\frac{RR'}{L} \right)'+\frac{\Lambda R^2 L}2 \right]\ 
  , \label{c1} 
  \\ {\mathcal H}_r^{\rm g} &= R' P_R - LP_L' \; ,\label{c2} \end{align} 
respectively (a prime denotes a derivative with respect to $r$). The 
boundary action $S_{\partial \Sigma}$ will be  
discussed below.  
 
The total action is the sum of \eqref{mega_action} and an action 
$S^{\rm d}$ 
describing the dust. The canonical formalism for the latter 
was developed in \cite{dust}. The dust action reads 
\begin{align} \label{matterH} S^{\rm d} 
  &= \int \mathrm{d}t \int_0^{\infty} \mathrm{d}r \left( P_{\tau} 
    \dot{\tau} 
- N {\mathcal H}^{\rm d} - 
    N^r {\mathcal H}_r^{\rm d} \right) \; ,\\ 
\intertext{where the contributions to the Hamiltonian and momentum 
    constraints are} \label{H_dust}{\mathcal H}^{\rm d} &= P_{\tau} 
  \sqrt{1+\frac{{\tau'}^2}{L^2}} \quad \mathrm{and} \quad {\mathcal 
    H}_r^{\rm d} = \tau' 
  P_{\tau} \; . \end{align} 
It is due to this relatively simple form of the dust action that exact 
quantum states can be found below. The full constraints are then the 
sum of the gravitational and the matter (dust) constraints. 
 
The general ansatz of the metric \eqref{metric} should, of course, 
correspond to the LTB metric \eqref{ltb-metric}. This leads to a 
couple of relations that are analogous to those derived in 
\cite{KMHV06}, except that ${\mathcal F}$ is now given by 
\eqref{mathcalF}. One can express, in particular, the variables $E$, 
$F$, and $\tau$, which completely characterize the collapse of the dust cloud, 
in terms of the canonical variables. 
 
What about the boundary action $S_{\partial \Sigma}$ in 
\eqref{mega_action}? In principle, we have to consider two 
boundaries. The first boundary is at the centre of symmetry for the dust cloud, 
the second either at spatial infinity or at another chosen end of the 
spatial hypersurfaces. As for the inner boundary, the effect of a 
positive cosmological constant compared to a vanishing or negative $\Lambda$ 
is there negligible, because its effects are felt only at cosmological 
distances. We can thus employ the boundary action used previously, 
cf. \cite{VGKS07}. It was found that the following term must be added 
to compensate a corresponding term that arises from variation of the 
action and partial integration: 
\be 
S_{\partial \Sigma}=\int_{\partial \Sigma}\D t\ N_0(t)M_0(t)\ , 
\ee 
where $N_0$ is the lapse function at $r=0$, and $M_0=F(0)/2G$ is the 
mass of the black hole at the centre. 
If this compensation were not made, one would conclude that a variation 
of $N_0$ leads to vanishing black-hole mass, which would certainly not 
be desirable. Such arguments were already used in the 
geometrodynamics of the pure Schwarzschild black hole \cite{Kuchar}. 
 
The choice of the second boundary is less obvious. Since the spacetime 
here is asymptotically de~Sitter, the spatial sections are no longer 
asymptotically flat. This leads basically to two options. One option 
is to choose a spacelike hypersurface that approaches the spacelike 
infinity in SdS spacetime. Here one must be careful and avoid $r$
to become timelike. This is, for example, achieved if we
employ coordinates such as the Painlev\'{e}--Gullstrand 
coordinates used in \cite{KL07} which are, in fact, close to the 
coordinates that we use below to simplify the constraints. Another 
option is to use the cosmological horizon as the second boundary. This 
would correspond to the choice frequently made in black-hole papers 
where the bifurcation sphere of the horizon is employed as the inner 
boundary, see, for example, \cite{LWH} and \cite{KL99}. It will also 
be our choice here.  
 
We thus consider the case where the cosmological horizon is  
the second boundary. The situation is then analogous to the cases considered in 
\cite{LWH} and \cite{KL99}. We choose $r=0$ 
to represent the centre of the cloud and let $r\to\infty$ at the 
cosmological horizon. This is possible if we smoothly match $r$ to 
the  tortoise coordinate $R_*$, defined by $\D R_*={\mathcal F}^{-1}\D r$, 
at the cosmological horizon.  
The time parameter $t$ is chosen equal to the Killing time $T$. 
Let us now turn to the 
fall-off conditions. As for $L$, we choose 
\begin{equation} 
L(t,r)=\frac{L_{0}(t)}{r^3}+O\left(\frac{1}{r^4}\right)\ . 
\label{fL} 
\end{equation} 
$L$ tends to zero near the cosmological horizon, where $r$ goes to 
infinity. This is required, since it is equal to the tortoise 
coordinate there. It is proportional to 
$1/r^3$, which guarantees that the distance to the horizon 
will be finite.  
As for $R$, we choose 
\begin{equation} 
R(t,r)=R_{\rm c} 
-\frac{R_1}{r^2}+\frac{R_2(t)}{r^3}+O\left(\frac{1}{r^4}\right)\ . 
\label{fR} 
\end{equation} 
For the remaining variables we impose 
\begin{equation} 
N(t,r)=\frac{N_0(t)}{r^2}+O\left(\frac{1}{r^3}\right)\ ,\quad 
N^r=O\left(\frac{1}{r}\right)\ ,  
\label{fN} 
\end{equation} 
and 
\begin{equation} 
P_L=O\left(\frac{1}{r}\right)\ ,\quad 
P_R=O\left(\frac{1}{r}\right) \label{fPl}\ .  
\end{equation} 
These fall-offs are consistent with the equations of motion and 
the constraint. They also guarantee that 
the boundary term at the cosmological horizon vanishes. 
 
One could now start with the quantization of \eqref{c1}, 
\eqref{c2}, and \eqref{matterH}.  
However, this turns out to be too complicated. One 
thus performs first a classical simplification of the constraints 
in order to render them manageable. This is achieved by 
introducing new variables and appropriate canonical 
transformations \cite{VWS01,KMHV06}. In this way, the variable 
$\Gamma\equiv F'$ and its momentum $P_{\Gamma}$ are introduced as 
new canonical variables; this choice is also convenient because 
one thereby absorbs a certain boundary term \cite{KMHV06}. Using 
then the momentum constraints to eliminate $P_{\Gamma}$ and 
squaring the ensuing Hamilton constraint, one arrives at the 
following final form of the constraints: The full Hamiltonian 
constraint reads \be \label{H-constraint} \mathcal{H} 
=P_{\tau}^2+\mathcal{F} P_R^2-\frac {\Gamma^2}{4G^2\mathcal{F}} 
\approx 0\ , \ee while the full momentum constraint reads \be 
\lb{Hr-constraint} \mathcal{H}_r = R' P_R -\Gamma P_\Gamma' + 
\tau' P_\tau \approx 0\ . \ee This is the form of the constraints 
suitable for quantization. Strictly speaking, re-defining the 
constraints is part of the very definition of quantization. There 
are many cases where the classical constraints are transformed 
into a manageable form by canonical transformations, see, for 
example, \cite{CJZ}. 
 
We note that because of the involved squaring the new Hamiltonian 
constraint \eqref{H-constraint} has acquired the dimension 
(mass/length)$^2$. We emphasize that the only differences compared to 
the earlier papers dealing with non-positive $\Lambda$ lie so far in the 
definition of $\mathcal{F}$ according to \eqref{mathcalF}. We also 
emphasize that through these manipulations the kinetic term in the 
Hamiltonian constraint is no longer hyperbolic, in contrast to the 
original form \eqref{c1}. More precisely, it is hyperbolic inside the 
black-hole horizon and outside the cosmological horizon, and it is 
elliptic between the horizons.\footnote{One might think that this
  change of sign is related to the fact that the metric in the
  interior of the horizon can become ``cosmological'', as it happens e.g.
  for the Schwarzschild black hole where the inside metric is of
  the Kantowski--Sachs form. This is, however, not the case. The
  equation ${\mathcal F}=0$ is a condition for an apparent horizon,
  and the sign change happens also for the non-static case where the
  interior is not ``cosmological''.}
A similar observation was made for the Reissner--Nordstr\"om 
black hole already in \cite{BK97}.\footnote{It is interesting to note 
that the Einstein equations for regular axisymmetric and stationary black 
holes surrounded by matter are also elliptic in the exterior and hyperbolic 
in the interior of the hole \cite{ansorg}.} This gives further support
for choosing a hypersurface that extends from one horizon to the
other. 
We finally note that the 
gravitational constant $G$ occurs in \eqref{H-constraint} in the same 
way as in the usual Hamiltonian constraint without matter; 
this may be of relevance for semiclassical 
approximation schemes in the quantum theory \cite{OUP}. 
 
The SdS spacetime is static only in between the two horizons. It thus 
makes sense to talk about a Killing time, $T$, only in this region. In 
previous papers it was shown that $T=2P_{\Gamma}$ and that 
\be 
\lb{killing} 
T=a \tau \pm \int \D R \ \frac{\sqrt{1-\mathcal{F} a^2}}{\mathcal{F}}\ , 
\ee 
where $a\equiv1/\sqrt{1+2E}$; the plus sign in \eqref{killing} holds 
for an expanding, the minus sign for a collapsing dust cloud. 
(This interpretation comes from the similarity of the coordinates 
$(\tau,R)$ with the Painlev\'{e}--Gullstrand coordinates for a 
Schwarzschild black hole \cite{MP01}.) It is 
straightforward to show that the same relation also holds in our case, 
taking into account the new definition for $\mathcal{F}$. Strictly 
speaking, one can talk about a Killing time only in the absence of the 
dust cloud, where the spacetime is of SdS form. However, as in the previous 
papers, one can approximately continue to interpret $T$ as a Killing 
time, because we  
consider the dust only as a small perturbation to the SdS spacetime. 

\section{Quantization}
 
We now turn to quantization. 
Applying the formal Dirac procedure, we replace the momenta by 
functional derivatives with respect to the corresponding configuration 
variables. From the Hamiltonian constraint \eqref{H-constraint} we get 
the Wheeler--DeWitt equation, 
\begin{align} \begin{split} \label{WDW_eq}\Bigg[- G^2 \hbar^2 \Bigg( 
    \frac{\delta^2}{\delta \tau(r)^2} & + \mathcal{F} \, 
    \frac{\delta^2}{\delta R(r)^2} + A(R,F) \, \delta(0) \, 
    \frac{\delta}{\delta R(r)} \\ & + B(R,F) \, \delta(0)^2 \Bigg) - 
    \frac{\Gamma^2}{4\mathcal{F}} \Bigg] 
    \Psi\left[\tau(r),R(r),\Gamma(r)\right] = 0 \; .\end{split} \end{align} 
Here, $A$ and $B$ are smooth functions of $R$ and $F$, which 
encapsulate the factor-ordering ambiguities. The factor-ordering 
problem is a fundamental problem that can only be dealt with after a 
suitable regularization procedure has been invoked. A general
treatment is beyond 
the scope of our paper. To implement the factor-ordering ambiguities 
at a formal level,  
we have introduced formal factors of $\delta(0)$ into the 
equation. The reason is that the equation can then be put onto a 
lattice and the continuum limit can be performed \cite{KMHV06}. This
corresponds to a particular regularization procedure that allows the
presentation of the solutions below.
 
 From \eqref{Hr-constraint} one gets the quantum momentum constraint 
\begin{align} \lb{qmc} \left[\tau' \frac{\delta}{\delta \tau(r)}  + R' 
    \frac{\delta}{\delta 
      R(r)} - \Gamma \left(\frac{\delta}{\delta \Gamma(r)}\right)' \right] 
  \Psi\left[\tau(r), R(r), \Gamma(r) \right] = 0 \; . \end{align} 
Both equations, \eqref{WDW_eq} and \eqref{qmc}, can be put on the 
lattice and a special class of exact solutions can be found 
\cite{KMHV06}. In the continuum limit, this class of solutions reads 
\bea 
\lb{Psi} 
& & \Psi\left[\tau(r), R(r), \Gamma(r) \right] 
=  
 \nonumber\\ 
& & \; \exp\left(\pm\frac{\I}{2G\hbar}\int\D r\ \Gamma\left[ a\tau+ 
    \int^R\D R \ \frac{\sqrt{1-a^2{\mathcal 
          F}}}{\mathcal{F}}\right]\right) \ . 
\eea 
(In the earlier papers, a more general solution with an unspecified 
constant $b$ was used \cite{KMHV06}; for simplicity, we set here $b=0$.) 
The momentum constraint \eqref{qmc} is solved by any state 
of the form $\exp\left(\I/(2G\hbar)\int\D r\ \Gamma W\right)$, where $W$ is a 
  smooth function of $R$ and $\tau$. The Wheeler--DeWitt equation, 
  however, is only solved for a particular factor ordering.  
Generalizing the treatment presented in \cite{KMHV06} to $\Lambda\neq 
0$, one must choose in \eqref{WDW_eq} $B=0$ and 
\be 
\lb{factorordering} 
A= \frac12\left(\frac{F}{R^2}-\frac{2\Lambda 
    R}{3}\right)\frac{2-a^2{\mathcal F}}{1-a^2{\mathcal F}}\ . 
\ee 
This means that $\Lambda$ does not only enter the definition of 
${\mathcal F}$, but also the chosen factor ordering. Exact 
solutions for the quantum states can only be found for this special 
$\Lambda$-dependent choice 
of $A$. This is certainly a weak point of our 
approach, because there should be a fixed factor ordering at the 
fundamental level. Nevertheless, this choice for $A$ gives us the 
opportunity to have exact solutions at our disposal. Moreover, at the 
semiclassical level of WKB states, the factor ordering is irrelevant, 
and the calculation of Hawking radiation below is an effect that 
occurs at the WKB level \cite{VKSW03}. In fact, one can use a 
semiclassical expansion scheme instead of the exact solutions to 
derive Hawking radiation and quantum gravitational corrections 
\cite{Banerjee}.  

\section{Hawking radiation}
 
Let us now turn to the calculation of Hawking radiation. The idea is 
to remain as close as possible to Hawking's original derivation in 
\cite{Hawking}, but to transfer this idea to solutions of the 
Wheeler--DeWitt equation and the momentum constraints, which are full 
quantum gravitational states. Because the Wheeler--DeWitt equation 
\eqref{WDW_eq} contains second derivatives with respect to the dust 
time $\tau$ (which is a consequence of squaring the Hamiltonian at the 
classical level), we can define states of positive and negative 
frequency with respect to $\tau$. This is crucial for our discussion.  
The situation is thus analogous to 
the Klein--Gordon equation, where positive frequency is defined by a 
minus sign in the phase, $\exp(-\I Et/\hbar)$, and correspondingly negative 
frequency by a positive sign. 
 
For the state with positive frequency we thus get from \eqref{Psi}, 
\bea 
\lb{Psiplus} 
& & \Psi^+\left[\tau(r), R(r), \Gamma(r) \right] 
 =  
 \nonumber\\ 
& & \; \exp\left(-\frac{\I}{2G\hbar}\int\D r\ \Gamma\left[ a\tau+ 
    \int^R\D R \ \frac{\sqrt{1-a^2{\mathcal 
          F}}}{\mathcal{F}}\right]\right) \ , 
\eea 
while for the state with negative energy we have 
\bea 
\lb{Psiminus} 
& & \Psi^-\left[\tau(r), R(r), \Gamma(r) \right] 
=  
 \nonumber\\ 
& & \; \exp\left(\frac{\I}{2G\hbar}\int\D r\ \Gamma\left[ a\tau+ 
    \int^R\D R \ \frac{\sqrt{1-a^2{\mathcal 
          F}}}{\mathcal{F}}\right]\right) \ . 
\eea 
In Hawking's calculation in \cite{Hawking}, the Bogolyubov coefficient 
$\beta$, which gives the negative-frequency part of the time-developed 
original state with positive frequency, plays a central role. 
 Here, the role of the quantum matter 
field in \cite{Hawking} is played by the dust. We thus define Hawking 
radiation for the black-hole case as the overlap between an ``outgoing 
dust state'' with 
negative energy and an ``ingoing dust state'' with positive energy. 
 Since the interpretation of these states is made with respect to an 
observer in the SdS spacetime using the Killing time $T$, we have to 
substitute the dust time by $T$ according to \eqref{killing}. 
Taking the corresponding signs in \eqref{killing} into account, we 
have then to calculate the overlap between 
\be 
\lb{Psiminuse} 
\Psi^-_{\rm e}= 
\exp\left(\frac{\I}{2G\hbar}\int\D r\ \Gamma T\right)\ , 
\ee 
which is actually independent of $R$, 
and 
\be 
\lb{Psiplusc} 
\Psi^+_{\rm c}=  
\exp\left(-\frac{\I}{2G\hbar}\int \D r\ \Gamma T 
  -\frac{\I}{G\hbar}\int\D 
  r\Gamma\int^R\D R\ \frac{\sqrt{1-a^2{\mathcal F}}}{\mathcal 
    F}\right)\ , 
\ee 
where the index e(c) denotes ``expanding'' (``collapsing''). 
 
For the cosmological horizon, the situation is just the opposite 
because the Hawking radiation is now incoming from the horizon. We 
thus have to calculate the overlap between the ``ingoing 
negative-energy state'' 
\be 
\lb{Psiminusc} 
\Psi^-_{\rm c}=  
 \exp\left(\frac{\I}{2G\hbar}\int \D r\ \Gamma T 
  +\frac{\I}{G\hbar}\int\D 
  r\Gamma\int^R\D R\ \frac{\sqrt{1-a^2{\mathcal F}}}{\mathcal 
    F}\right) 
\ee 
and the ``outgoing positive-energy state'' 
\be 
\lb{Psipluse} 
\Psi^+_{\rm e}= 
\exp\left(-\frac{\I}{2G\hbar}\int\D r\ \Gamma T\right)\ . 
\ee 
 
In order to calculate the overlap between the quantum states 
\eqref{Psiminuse} and \eqref{Psiplusc}, or between \eqref{Psiminusc} 
and \eqref{Psipluse}, 
we need a Hilbert space. As 
far as full quantum gravity is concerned, this is an open issue 
\cite{OUP}. In the present case, however, a natural candidate is 
present \cite{VKSW03,KMHSV07}. We can choose the standard 
Schr\"odinger inner product with respect to constant time and $R$ as 
the integration variable. The integration range is supposed to run 
from the black-hole horizon, $R_{\rm h}$, to the cosmological horizon, 
$R_{\rm c}$. This is further motivated by the fact that our 
hypersurface lies between the horizons; as we have already noted, 
in this region the kinetic 
term in \eqref{H-constraint} is elliptic. 
 
Addressing first the cosmological horizon, we have for the desired 
overlap between ingoing negative-energy and outgoing 
positive-energy states the expression 
\be 
\lb{innerproduct}\langle\Psi^-_{\rm c}\vert\Psi^+_{\rm e}\rangle\equiv 
\int_{R_{\rm h}}^{R_{\rm c}}\D R\ \sqrt{g_{RR}}\Psi^{-*}_{\rm c} 
\Psi^+_{\rm e}\ . 
\ee 
Here, $g_{RR}$ is the $RR$-component of the DeWitt metric, as it can 
be read off \eqref{H-constraint} where its inverse is the prefactor of 
the term $P_R^2$, and we thus have $g_{RR}={\mathcal 
  F}^{-1}$.\footnote{Note that this measure is not the measure that 
  would render the Wheeler--DeWitt equation \eqref{WDW_eq} with the 
  factor ordering \eqref{factorordering} Hermitean \cite{VGKS07}. Our 
  choice of the measure in \eqref{innerproduct} would make the 
  $R$-part of \eqref{WDW_eq} Hermitean if the Laplace--Beltrami factor 
  ordering were used.} 
 
There is, however, one further point to consider. Since we are 
interested in an observer far away from the dust cloud, for 
whom the Killing time $T$ according to \eqref{killing} is the 
appropriate time coordinate, we have already rewritten the states in 
terms of $T$. Consequently, as in \cite{VKSW03,KMHSV07}, 
we have to evaluate also the component $g_{RR}$ in the new coordinate 
system $(T,R)$ instead of $(\tau,R)$. Using \eqref{killing}, one gets 
$\sqrt{g_{RR}}=(a{\mathcal F})^{-1}$. This is the expression 
to be used in the calculation of the Bogolyubov coefficient. 
 
For the concrete calculation we shall write the full states as a 
product of single-shell states where the radial variable $r$ is 
assumed to consist of discrete points separated by a distance 
$\sigma$. (The continuum limit is obtained for $\sigma\to 0$.) As in 
\cite{KMHSV07}, the Bogolyubov coefficient $\beta$ is calculated for 
each shell separately. In the discrete case, we replace $\Gamma$ by 
the dimensionless variable $2\omega$ and indicate the dependence on 
$\omega$ by an index. (The factor $2$ is motivated by 
the fact that $\Gamma=2M'$.) 
We omit the shell index and write the 
corresponding wave functions as $\psi_{\omega}(T,R)$. From 
\eqref{innerproduct} we then define $\beta$ to read 
\be 
\lb{beta} 
\beta_{\omega\omega'}\equiv \int_{R_{\rm h}}^{R_{\rm c}}\D R\ 
\sqrt{g_{RR}}\Psi^{-*}_{{\rm c}\omega} \Psi^+_{{\rm e}\omega'}\ . 
\ee 
In the earlier papers \cite{KMHSV07} and \cite{VGKS07} we have chosen 
a particular normalization for $\beta$, which was motivated by the 
normalization of states for the Klein--Gordon equation. Since the full 
normalization of solutions to the Wheeler--DeWitt equation is anyway 
not known, we shall leave the question of normalization open here and 
define $\beta$ directly by \eqref{beta}. 
 
Employing now the one-shell contributions of \eqref{Psiminusc} and 
\eqref{Psipluse} and inserting them into \eqref{beta}, one gets the 
following expression: 
\bea 
\lb{betaintegral} 
\beta_{\omega\omega'} &=& 
\sqrt{1+2E}\exp\left(-\frac{\I\sigma 
    T}{G\hbar}[\omega+\omega']\right)\times\nonumber\\ 
& & \; \int_{R_{\rm h}}^{R_{\rm c}}\D R\ {\mathcal 
F}^{-1}\exp\left(-\frac{2\I\sigma\omega}{G\hbar}\int^R\D 
R\frac{\sqrt{1-a^2{\mathcal F}}}{\mathcal F}\right)\ . 
\eea 
We shall first calculate the contribution to 
$\beta_{\omega\omega'}$ from the cosmological horizon. This will be 
achieved through an appropriate near-horizon approximation,
making the standard assumption that the integral is dominated 
by the close vicinity of the horizon. We start 
by writing ${\mathcal F}^{-1}$ in terms of partial fractions: 
\bea 
\label{partial} 
\frac {1}{\cF} &=&\frac{1}{1-\frac FR -\frac{\Lambda R^2}{3}}\nonumber \\ 
&=& R \left[\frac{A}{R-R_{\rm c}}+\frac{B}{R-R_{\rm 
      h}}+\frac{C}{R-R_{\rm n}} \right] . 
\eea 
Now we introduce the variable $s$ by $ls=R_{\rm c}-R$ so that near the 
cosmological horizon $\vert s \vert\ll 1$. Here, $l\equiv 1/\sqrt{\Lambda}$.  
 
First we consider the term 
\bea 
\int^R \md R \frac{\sqrt{1-a^2{\mathcal F}}}{\mathcal{F}}&=&\int^s 
-l\md s 
(R_{\rm c}-ls)\left[-\frac{A}{ls}+\frac{B}{R_{\rm c}-R_{\rm 
      h}-ls}+\frac{C}{R_{\rm c}-R_{\rm n}-ls}  
\right]  \nonumber\\  
 & & \times \sqrt{1-\frac{a^2}{R_{\rm c}-ls} 
   \left(\frac{1}{-\frac{A}{ls}+\frac{B}{R_{\rm c}-R_{\rm 
         h}-ls}+\frac{C}{R_{\rm c}-R_{\rm n}-ls}}\right)}. 
\eea 
After some calculation, neglecting terms with a positive power in $s$, we get: 
\bea 
\label{x}
 \int^R \md R \frac{\sqrt{1-a^2{\mathcal F}}}{\mathcal{F}}\approx
\int^s \md s \sqrt{\mbox{const}-\frac 1s \xi + \frac{1}{s^2} 
  A^2R^2_{\rm c}}\ , 
\eea 
with $\xi=2A^2lR_{\rm c}+\frac{2ABR_{\rm c}^2l}{R_{\rm c}-R_{\rm 
    h}}+\frac{2ACR_{\rm c}^2l}{R_{\rm c}-R_{\rm n}}-a^2AlR_{\rm c}$.  
Expanding \eqref{x} for $|s|\ll 1$ yields 
\bea 
\int^s \md s \sqrt{\mbox{const}-\frac 1s \xi + \frac{1}{s^2} A^2R^2_c} 
&\approx &  
\int^s \md s \frac 1s \left[AR_c-\frac{\xi s}{2AR_c}+\cO(s^2) \right] 
\nonumber \\  
&\approx & AR_c \ln|s|-\frac{\xi s}{2AR_c}. 
\eea 
Inserting this in our integral \eqref{betaintegral} and using
\bdm
{\mathcal F}^{-1}\approx \frac{AR}{R-R_{\rm c}}\approx -\frac{AR_{\rm
    c}}{ls}\ ,
\edm
 we obtain 
\bea 
\label{beta2} 
 \beta_{\omega\omega'}&=& - \sqrt{1+2E} 
 \exp{\left(-\I\frac{\sigma 
  T}{G\hbar}(\omega+\omega')\right)} AR_{\rm c} \nonumber\\ 
& & \times \int_{0}^{(R_{\rm c}-R_{\rm h})/l} \md s 
s^{{-2\I\sigma\omega AR_{\rm c}}/{G\hbar}-1}  
\exp{\left[ \frac{\I\xi}{AR_{\rm c}} \frac{\sigma\omega s}{G\hbar}\right]} \ . 
\eea 
To evaluate this integral, we use the formula \cite{gradshteyn}
\begin{displaymath} 
\int_0^{\infty}\md x\ x^{\nu-1}e^{-(p+\I q)x}=\Gamma(\nu)(p^2+q^2)^{-\nu/2} 
  e^{-\I\nu\mathrm{arctan}(q/p)} \ ,
\end{displaymath} 
which is, in particular, applicable to the case $p=0$ and $0<{\rm
  Re}\nu<1$.
(We insert a small positive value for ${\rm Re}\nu$, which we let go
to zero after the integration.) Then,
\bea 
& & \beta_{\omega\omega'}=-\sqrt{1+2E} 
\exp\left(-\I\frac{\sigma 
  T}{G\hbar}(\omega+\omega')\right)A R_{\rm c} 
\times \nonumber\\ 
& & \ \Gamma\left(-\frac{2\I\sigma\omega AR_{\rm c}}{G\hbar}\right) 
   \left(-\frac{\sigma\omega \xi}{G\hbar AR_{\rm c}}\right)^{2\I\sigma\omega 
AR_{\rm c}/G\hbar}e^{\pi\sigma\omega AR_{\rm c}/G\hbar}\ . 
\eea 
Using
\bdm 
\Gamma\left(-\I y\right)\Gamma\left(\I y\right)=\frac{\pi}{y\sinh \pi
  y}
\edm 
(with real $y$), we get 
\be 
\vert\beta_{\omega\omega'}\vert^2 = -\frac{\pi 
   G\hbar AR_{\rm c}(1+2E)}{\sigma\omega} 
\frac{1}{e^{-4\pi \sigma\omega 
    AR_c/G\hbar} - 1}\ 
\ee 
for the absolute square of $\beta$. (In spite of the minus sign, this
is positive because $A<0$, see below.)
Next we want to calculate the particle creation rate, which means we 
have to evaluate the expression $\Sigma_k 
\beta_{ik}\beta_{ik}^*$. Here this corresponds to the integral 
$\int_{0}^\infty \md \omega'|\beta_{\omega\omega'}|^2$. This integral diverges and hence needs to be suitably regulated. We introduce a decay factor $e^{-b\omega'}$ and carry out the integration.    
For the ``in'' particle number operator we then obtain 
\be 
\lb{N11} 
\langle\mathrm{in}\vert\hat{N}_{\mathrm{in}}\vert\mathrm{in}\rangle= 
-\frac{\pi 
   G\hbar AR_{\rm c}(1+2E)}{b\sigma\omega}\frac{1}  
{e^{-4\pi \sigma\omega AR_c/G\hbar} - 1}\ . 
\ee 
Replacing $\sigma\omega$ by $G\Delta\epsilon$, where 
$\Delta\epsilon$ is the energy of a shell, we arrive at the final result 
\be 
\lb{result} 
\langle\mathrm{in}\vert\hat{N}_{\mathrm{in}}\vert\mathrm{in}\rangle= 
-\frac{\pi AR_{\rm c} \hbar(1+2E)}{b\Delta\epsilon  } 
\frac{1}{e^{-4\pi AR_c\Delta\epsilon /\hbar} -1}\ , 
\ee 
from which we read off 
\be 
k_{\mathrm{B}}T_{\rm c}=-\frac{\hbar}{4\pi AR_{\rm c}} \ . 
\ee 
How can one interpret the prefactor in \eqref{result}? It should 
give the greybody factor, but its interpretation is complicated by the 
fact that the full normalization of our wave functionals is open. 
We thus do not address it any further. 
 
Using \eqref{partial}, one can easily determine 
\bdm
A=-\frac{3}{\Lambda 
  (R_{\rm c}-R_{\rm h})(R_{\rm c}-R_{\rm n})} 
\edm
and thus finds 
\be 
\lb{temperature_c} 
k_{\mathrm{B}}T_{\rm c}=\frac{\hbar \Lambda (R_{\rm c}-R_{\rm 
    h})(R_{\rm c}-R_{\rm n})}{12\pi R_{\rm c}} \ . 
\ee 
This result coincides with Equation (2.15b) in \cite{GH77}.  
 
Interestingly, our calculation yields a Planck spectrum with a 
temperature that would be measured by a fictitious observer with 
$g_{TT}=1$. While such an observer could be easily realized in the 
Schwarzschild case (he would be situated at spatial infinity), this is 
not possible here. For the SdS case, the observer should be located 
between the horizons. Following \cite{BH96}, we shall consider the 
preferred observer who is stationary at a position $R_{\rm o}$ where 
the black-hole attraction and the cosmological expansion cancel each 
other; such an observer is thus unaccelerated and moves on a 
geodesic. We have\footnote{For the above example of the supermassive 
  black hole in the quasar OJ287, one gets $R_{\rm o}\approx 260$ kpc, 
which corresponds to roughly five times our distance to the Large 
Magellanic Cloud!} 
\be 
R_{\rm o}=\left(\frac{3GM}{\Lambda}\right)^{1/3}\ , 
\ee 
which leads to the modified Hawking temperature 
\be 
\lb{temperature_cmodified} 
k_{\mathrm{B}}T_{\rm c}=\frac{1}{\sqrt{1-(9G^2M^2\Lambda)^{1/3}}} 
\frac{\hbar \Lambda (R_{\rm c}-R_{\rm 
    h})(R_{\rm c}-R_{\rm n})}{12\pi R_{\rm c}} \ . 
\ee 
This also agrees with the corresponding expression for the temperature 
given in Equation (17) in \cite{Stotyn} where the Hawking temperature is 
calculated from a tunnelling picture. 
 
A completely analogous calculation yields the Hawking temperature for 
the black-hole horizon. In fact, the corresponding overlap between 
$\Psi_{\rm e}^-$ and $\Psi_{\rm c}^+$ leads to the same expression 
\eqref{beta} for $\beta$. In contrast to above, however, the 
near-horizon approximation is now performed for the black-hole 
horizon. Now, the coefficient $B$ in \eqref{partial} enters. It is 
given by 
\bdm 
B=-\frac{3}{\Lambda 
  (R_{\rm h}-R_{\rm c})(R_{\rm h}-R_{\rm n})}\ , 
\edm 
which follows from the expression for $A$ by interchanging $R_{\rm c}$ 
and $R_{\rm h}$. 
One thereby arrives at the expression 
\be 
\lb{temperature_h} 
k_{\mathrm{B}}T_{\rm h}=\frac{\hbar \Lambda (R_{\rm c}-R_{\rm 
    h})(R_{\rm h}-R_{\rm n})}{12\pi R_{\rm h}} \ , 
\ee 
which is equal to Eq. (2.15a) in \cite{GH77}. This expression together 
with \eqref{temperature_c} are the main results of our paper, because 
they have been derived from candidates for exact quantum gravitational 
states.  
The temperature \eqref{temperature_h} will be modified by the same factor as in 
\eqref{temperature_cmodified} if an observer is considered who follows 
the geodesic motion at $R_{\rm o}$.  
 
The two temperatures \eqref{temperature_c} and \eqref{temperature_h} 
arise from the same expression \eqref{beta} by two different 
near-horizon approximations. An exact evaluation of \eqref{beta} would 
yield the sum of both temperatures plus interference terms between the 
two types of Hawking radiation. In the extremal case of the Nariai 
metric, both temperatures are zero. Nevertheless, an exact evaluation 
of \eqref{beta} could yield a small non-zero contribution which would 
be a genuine quantum gravitational effect. 
 
To summarize, we have shown that solutions to the Wheeler--DeWitt 
equation and the diffeomorphism constraints contain information about 
the two Hawking temperatures from the black-hole and the cosmological 
horizon. An interesting open question is the calculation of the
entropy for the Schwarzschild--de~Sitter case through a counting of
microstates. Such a derivation was presented for the BTZ black hole
\cite{VGKSW} and the AdS black hole \cite{VW09}. Whether it is also
possible here is left for future publications.
\\ 
\\ 
{\bf Acknowledgements:} 
We thank Rabin Banerjee, Max D\"orner, T. P. Singh, Rakesh Tibrewala,
and Cenalo Vaz for helpful discussions and critical comments.  
A.F. thanks the Tata Institute of 
Fundamental Research, Mumbai, for kind hospitality during her stay. 

 
\end{document}